\documentclass[twocolumn]{elsart}

\usepackage{graphicx}
\usepackage{amssymb}
\usepackage{amsmath}
\usepackage{upgreek}
\usepackage{times}
\usepackage[american]{babel} 
\usepackage{umlaut}
\usepackage{xspace}         %

\journal{Nucl. Instr. Meth. in Phys. Res. Sec. A}

\newcommand{\aman}{{\sc Aman\-da}\xspace}

\newcommand{\ice}{{\sc Ice\-Cube}\xspace}
\newcommand{\st}{{\sc String-18}\xspace}

\newcommand{\cer}{Che\-ren\-kov\xspace}
\newcommand{\dc}{{\sc DOMCom}\xspace}

\def\be{\par\nobreak\noindent\begin{equation}}
\def\ee{\end{equation}\par\nobreak\noindent}
\def\beqn{\par\nobreak\noindent\begin{eqnarray}}
\def\eeqn{\end{eqnarray}\par\nobreak\noindent}

\begin{document}
\bibliographystyle{h-elsevier2}

\begin{frontmatter}
\title{%
The \textbf{\textsc{IceCube}}
prototype string in \textbf{\textsc{Amanda}}}
{\large The \aman Collaboration} %
\author[4]{M.~Ackermann},
\author[11]{J.~Ahrens},
\author[1]{X.~Bai},
\author[20]{M. Bartelt},
\author[10]{S.W.~Barwick},
\author[9]{R.~Bay},
\author[11]{T.~Becka},
\author[20]{J.K.~Becker},
\author[2]{K.-H.~Becker},
\author[4]{E.~Bernardini},
\author[3]{D.~Bertrand},
\author[4]{D.J.~Boersma},
\author[4]{S.~B\"oser}, 
\author[17]{O.~Botner},
\author[17]{A.~Bouchta}, 
\author[3]{O.~Bouhali}, 
\author[15]{J.~Braun}
\author[18]{C.~Burgess},
\author[18]{T.~Burgess},
\author[13]{T.~Castermans},
\author[7]{W.~Chinowsky}, 
\author[7]{D.~Chirkin}, 
\author[17]{J.~Conrad}, 
\author[15]{J.~Cooley},
\author[8]{D.F.~Cowen}, 
\author[17]{A.~Davour}, 
\author[19]{C.~De~Clercq},
\author[12]{T.~DeYoung},
\author[15]{P.~Desiati}, 
\author[18]{P.~Ekstr\"om}, 
\author[11]{T.~Feser}, 
\author[4]{M.~Gaug},
\author[1]{T.K.~Gaisser}, 
\author[15]{R.~Ganugapati}, 
\author[2]{H.~Geenen}, 
\author[10]{L.~Gerhardt}, 
\author[7]{A.~Goldschmidt},
\author[20]{A.~Gro\ss},
\author[17]{A.~Hallgren}, 
\author[15]{F.~Halzen}, 
\author[15]{K.~Hanson}, 
\author[9]{R.~Hardtke}, 
\author[2]{T.~Harenberg},
\author[1]{T.~Hauschildt}, 
\author[2]{K.~Helbing\corauthref{cor1}},
\ead{\textrm{helbing@uni-wuppertal.de}}
\corauth[cor1]{Corresponding author: Klaus Helbing. 
Address: Fachbereich C -- Physik, Gau{\ss}str. 20, Bergische Universit\"at Wuppertal , D-42119
Wuppertal, Germany;
Phone: +49 202 439 2829
Fax: +49 202 439 2662 }
\author[11]{M.~Hellwig}, 
\author[13]{P.~Herquet}, 
\author[15]{G.C.~Hill}, 
\author[19]{D.~Hubert}, 
\author[15]{B.~Hughey}, 
\author[18]{P.O.~Hulth}, 
\author[18]{K.~Hultqvist},
\author[18]{S.~Hundertmark}, 
\author[7]{J.~Jacobsen}, 
\author[2]{K.H.~Kampert},
\author[15]{A.~Karle}, 
\author[15]{J.L.~Kelley},
\author[8]{M.~Kestel},
\author[13]{G.~Kohnen}, 
\author[11]{L.~K\"opke}, 
\author[4]{M.~Kowalski}, 
\author[15]{M.~Krasberg},
\author[10]{K.~Kuehn}, 
\author[4]{H.~Leich}, 
\author[4]{M.~Leuthold}, 
\author[5]{I.~Liubarsky},
\author[7]{J.~Ludvig\thanksref{sunny}}\thanks[sunny]{now at Stanford
Research Systems, Sunnyvale, CA 94089, USA},
\author[17]{J.~Lundberg},
\author[16]{J.~Madsen},
\author[17]{P.~Marciniewski}, 
\author[7]{H.S.~Matis}, 
\author[7]{C.P.~McParland}, 
\author[20]{T.~Messarius}, 
\author[18]{Y.~Minaeva}, 
\author[9]{P.~Mio\v{c}inovi\'c}, 
\author[15]{R.~Morse}, 
\author[20]{K.S.~M\"unich},
\author[4]{R.~Nahnhauer}, 
\author[10]{J.~Nam},
\author[11]{T.~Neunh\"offer}, 
\author[1]{P.~Niessen}, 
\author[7]{D.R.~Nygren}, 
\author[15]{H.~\"Ogelman}, 
\author[19]{Ph.~Olbrechts}, 
\author[17]{C.~P\'erez~de~los~Heros}, 
\author[6]{A.C.~Pohl}, 
\author[9]{R.~Porrata},
\author[9]{P.B.~Price}, 
\author[7]{G.T.~Przybylski}, 
\author[15]{K.~Rawlins}, 
\author[4]{E.~Resconi}, 
\author[20]{W.~Rhode}, 
\author[13]{M.~Ribordy}, 
\author[15]{S.~Richter}, 
\author[2]{S.~Robbins},
\author[18]{J.~Rodr\'\i guez~Martino}, 
\author[11]{H.-G.~Sander}, 
\author[4]{S.~Schlenstedt}, 
\author[15]{D.~Schneider}, 
\author[15]{R.~Schwarz}, 
\author[10]{A.~Silvestri}, 
\author[9]{M.~Solarz}, 
\author[7]{J.~Sopher}
\author[16]{G.M.~Spiczak}, 
\author[4]{C.~Spiering}, 
\author[15]{M.~Stamatikos}, 
\author[15]{D.~Steele}, 
\author[4]{P.~Steffen}, 
\author[7]{R.G.~Stokstad}, 
\author[4]{K.-H.~Sulanke}, 
\author[14]{I.~Taboada}
\author[18]{L.~Thollander}, 
\author[1]{S.~Tilav}, 
\author[20]{W.~Wagner}, 
\author[18]{C.~Walck}, 
\author[4]{M.~Walter},
\author[15]{Y.-R.~Wang}, 
\author[15]{C.~Wendt},
\author[2]{C.H.~Wiebusch},
\author[4]{R.~Wischnewski}, 
\author[4]{H.~Wissing}, 
\author[9]{K.~Woschnagg}, 
\author[10]{G.~Yodh}
\small
\address[1]{Bartol Research Institute, University of Delaware, Newark, DE 19716, USA}
\address[2]{Fachbereich C -- Physik, BU Wuppertal, D-42119 Wuppertal, Germany}
\address[3]{Universit\'e Libre de Bruxelles, Science Faculty,
Brussels, Belgium}
\address[4]{DESY-Zeuthen, D-15738 Zeuthen, Germany}
\address[5]{Blackett Laboratory, Imperial College, London SW7 2BW, UK}
\address[6]{Dept.~of Chemistry and Biomedical Sciences, University of Kalmar, S-39182~Kalmar,~Sweden}
\address[7]{Lawrence Berkeley National Laboratory, Berkeley, CA 94720, USA}
\address[8]{Dept.~of~Physics,~Pennsylvania~State~University,~University~Park,~PA~16802,~USA}
\address[9]{Dept. of Physics, University of California, Berkeley, CA 94720, USA}
\address[10]{Dept. of Physics and Astronomy, Univ. of California,
Irvine, CA 92697, USA}
\address[11]{Institute of Physics, University of Mainz, D-55099 Mainz, Germany}
\address[12]{Dept. of Physics, University of Maryland, College Park, MD 20742, USA}
\address[13]{University of Mons-Hainaut, 7000 Mons, Belgium}
\address[14]{Dept. de F\'{\i}sica, Universidad Sim\'on Bol\'{\i}var,
Caracas, 1080, Venezuela}
\address[15]{Dept. of Physics, University of Wisconsin, Madison, WI
53706, USA}
\address[16]{Physics Dept., University of Wisconsin, River Falls, WI
54022, USA}
\address[17]{Div. of High Energy Physics, Uppsala University, S-75121
Uppsala, Sweden}
\address[18]{Dept. of Physics, Stockholm University, SE-10691 Stockholm, Sweden}
\address[19]{Vrije Universiteit Brussel, Dienst ELEM, B-1050 Brussels,
Belgium}
\address[20]{Institute of Physics, University of Dortmund, D-44221
Dortmund, Germany}

\begin{abstract}
The {\bf A}ntarctic {\bf M}uon {\bf A}nd {\bf N}eutrino {\bf D}etector {\bf
A}rray (\aman) is a high-energy neutrino telescope.   
It is a lattice of optical modules (OM) installed in the
clear ice below the South Pole Station. Each OM contains a
photomultiplier tube (PMT) that
detects photons of \cer light generated in the
ice by muons and electrons. 
\ice is a cubic-kilometer-sized expansion of \aman currently being
built at the South Pole. 
In \ice the PMT signals are digitized already in the optical
modules and transmitted to the surface.
A prototype string of 41 OMs equipped with this new all-digital
technology was deployed in the \aman array in the year 2000. 
In this paper we describe the technology and demonstrate that this
string serves as a proof of concept for the \ice array. Our
investigations show that the OM timing accuracy is 5~ns. 
Atmospheric muons are detected in excellent agreement with
expectations with respect to both angular distribution and absolute rate.
\end{abstract}

\end{frontmatter}

\section{Introduction}
\label{sec:intro}
The {\bf A}ntarctic {\bf M}uon {\bf A}nd {\bf N}eutrino {\bf D}etector {\bf
A}rray (\aman)~\cite{Amanda:2001} at the
geographic South Pole is a lattice of
photomultiplier tubes (PMTs) each enclosed in a
transparent pressure sphere to comprise an optical module (OM). The
OMs are buried in the polar ice at depths between 1500\,m and 2000\,m.  
The primary goal of \aman is to
discover astrophysical sources of high energy neutrinos. 
High-energy muon neutrinos penetrating the earth
from the Northern Hemisphere are identified by the secondary muons
produced in charged current neutrino-nucleon interactions near or
within the detector.  
\aman was the first neutrino telescope with an effective area in
excess of 10000~m$^2$ -- with over 600 optical modules in place.

Large as it is, \aman is only capable of seeing the
brighter (and closer) sources of neutrinos. \ice will be an array
of 4,800 optical modules within a cubic kilometer of 
ice~\cite{PDD:2001}. Frozen into vertical holes 2.4~km deep, drilled by hot
water, the optical modules will lie between 1450~m and 2450~m below
the surface. An instrument of this size is
suited to study neutrinos from distant astrophysical
sources~\cite{Ahrens:2003ix}. 

The optical modules in \aman and \ice 
are arranged like
strings of beads connected electrically and mechanically to long
cables
leading up to a control room
at the surface. 
The PMT within the OM  detects individual photons of
\cer light generated in the ice by muons and
electrons moving with velocities near the speed of light. 
Signal events consist primarily of up-going muons produced in neutrino
interactions in the bedrock of Antarctica or in the ice. 
In addition, the detector will identify electromagnetic and hadronic showers
(``cascades'') from interactions of  $\nu_e$ and $\nu_\tau$ inside the
detector volume provided they are sufficiently energetic. 
Background events are mainly downward-going muons from
cosmic ray interactions in the atmosphere above the detector. For \ice,
the background will be monitored for calibration purposes by the
\textsc{IceTop} air shower array~\cite{PDD:2001} covering a 1~km$^2$ area at the ice
surface right above the in-ice OMs of the \ice detector. This
background will also be used as a test beam for commissioning the
detector. 

Unlike \aman, \ice PMT signals are digitized
directly in the optical module itself and then transmitted digitally to the
surface. This is done such that the arrival time of a photon at an OM
can be determined to within a few nanoseconds. The electronics at the
surface determines when an event has occurred (e.g., that a muon
traversed or passed near the array) and records the information for
subsequent event reconstruction and analysis. 

In the following we will discuss the performance obtained with a
string of 41 Digital Optical Module (DOM) prototypes 12~m apart
deployed in January 2000 by the \aman Collaboration as the
18$^\text{th}$ string (\st) of the \aman array. 

\section{Digital technology}

The DOMs of \st were intended to
demonstrate the advantages and the
feasibility of a purely digital technical approach, in which PMT
signals are captured and digitized locally, within the optical
module. The digital data are transmitted to the surface using the
twisted pair copper conductors that also bring power and control
signals to the DOMs. 
We will discuss the design, data
taking and data analysis of \st. 

\subsection{Concept}
Very large high-energy cosmic neutrino detectors require photomultiplier
tubes to be located far from the signal processing center. 
The completed \aman
detector represents an evolution of signal transmission technologies from
coaxial cable to twisted copper pair to optical fiber. In all these
cases, the analog signals are brought to the surface, where they are
digitized and processed. 

The longer cable lengths, the much larger number of PMTs of \ice and
the requirements of a larger dynamic range for PMT pulses and improved
time resolution, as
compared to \aman, have led to the development of a data-acquisition
technology with the following features: 
\begin{enumerate}
\item Robust copper cable and connectors between the surface
and the modules at depth.
\item Digitization and time-stamping of signals that are
unattenuated and undispersed. 
\item Calibration methods (particularly for timing)
that are appropriate for a very large number of optical modules. 
\end{enumerate}

The PMT anode signal is digitized and time-stamped already in the optical
module. Waveform digitization, in which all the information
in a PMT anode signal is captured, is incorporated. The time calibration
procedure is both accurate and automatic. 

There are three fundamental elements to the data acquisition
architecture of \ice and \st:
\begin{itemize}
\item The Digital Optical Module (DOM), which
captures the signals induced by physical processes and preserves the
information quality through immediate conversion to a digital format
using an innovative ASIC, the Analog Transient Waveform Digitizer
(ATWD~\cite{Kleinfelder:2003}, see Sec.~\ref{sec:info}).
\item A Network, which connects the highly dispersed array of
optical modules to the surface data acquisition system (DAQ) and
provides power to them.
\item  A Surface DAQ, which maintains a master clock time-base and
provides all messaging, data flow, filtering, monitoring, calibration,
and control functions. 
\end{itemize} 
\begin{figure*}
\includegraphics*[bb=73 330 571 731,width=\textwidth]{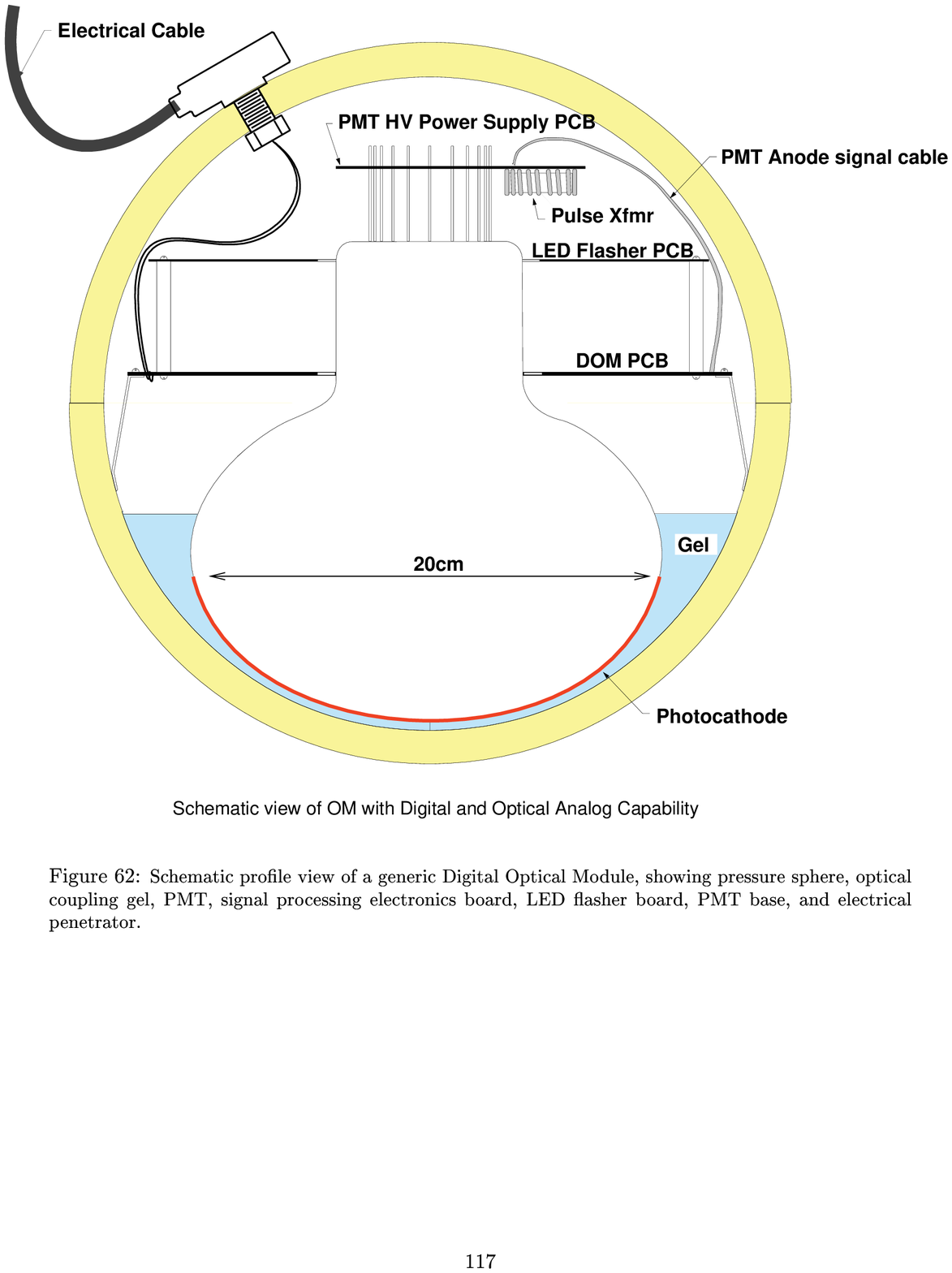}
\caption{Schematic profile view of a \st Digital Optical Module, 
showing pressure sphere, optical coupling gel, photomultiplier,
digital signal processing board (DOM PCB), LED flasher board, PMT
base, and electrical penetrator.} 
\label{fig:DOM}
\end{figure*}

\subsection{Digital optical module (DOM)}
The basic elements of the DOM are the PMT for detection of \cer
light, a printed circuit board (PCB) for processing signals, a HV
generator for the PMT, and an optical beacon consisting of LEDs
for calibration 
purposes all within the glass pressure housing. The general physical layout
of a DOM is shown in Fig.~\ref{fig:DOM}.

\subsubsection{Pressure Housing}
The spherical glass pressure housings are standard items used in
numerous oceanographic and maritime applications. 
The most important qualities are mechanical reliability, cost, optical
transmission, low radioactivity, and low scintillation of the glass. 
For \st ``Benthospheres'' are used, a trade-name of Benthos, Inc.

\subsubsection{Photomultiplier}
The optical sensor of the DOM is a large (8 inch diameter) hemispherical
14-stage photomultiplier tube, 
the Hamamatsu R5912-02 with a gain of $4 \cdot 10^8$ at the
typical operational voltage of 1400~V. These large PMTs offer good
time-resolution, as 
indicated by a transit-time-spread (TTS) for single photoelectron
pulses of about 2.5~ns rms. Despite their large photocathode
area, these PMTs alone\footnote{The dominant source of noise of the
integrated OM is the glass of the pressure sphere which leads to a
total of about 1~kHz noise pulse rate.} generate only  300~Hz or less of dark
noise pulses at temperatures below freezing and 1400~V. 
The photocathode sensitivity extends well into the UV. The
reception of \cer photons is limited by
the optical transmission of the glass pressure sphere at 350 nm. 
The PMT HV generator is a multi-stage voltage multiplier with an
oscillator running at a few tens of kHz. It converts the
electrical power transmitted via copper cable down into the ice to a
high voltage for the PMT.

\begin{figure}
\includegraphics[angle=-90,width=\columnwidth]{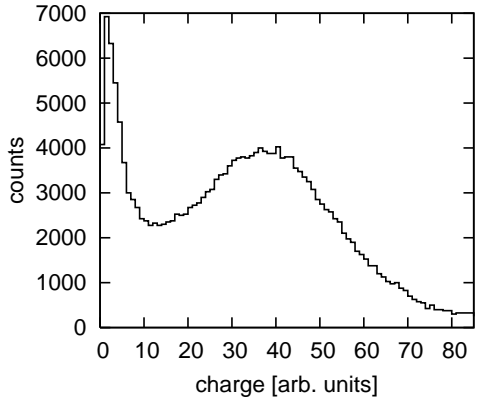}
\caption{Measured integral charge distribution of DOM~\#1. Most of
the histogram is single photoelectron noise pulses.}
\label{fig:SPE}
\end{figure}

For each photon that produces a single photoelectron, the
PMT produces pulses that have characteristic rise (fall) times of
7~(11)~ns. Due to the stochastic 
nature of the PMT's electron multiplication process, single
photoelectron pulses display 
significant variations in pulse shape and amplitude. The measured
integrated charge distribution of the deepest DOM (DOM~\#1) is shown in
Fig.~\ref{fig:SPE}.  The peak to valley ratio is very good for an
8~inch photomultiplier. 

\subsubsection{Optical beacon}
\label{sec:beacon}
Each of the \st DOMs is equipped with 6~GaN LEDs, which emit
predominantly in the near-UV at 370 nm. The luminous intensity and the
pulsing rate may be varied over a wide range under software
control. At their brightest, these beacons can be seen by OMs up to
distances of 200~m. The pulse width is  5~ns. The LEDs are broad angular
emitters, spaced at 60$^\circ$  around a vertical axis, and are canted
over by 45$^\circ$ to produce a roughly hemispherical source. 

\subsubsection{Signal processing circuitry}
\begin{figure*}
\includegraphics[width=\textwidth]{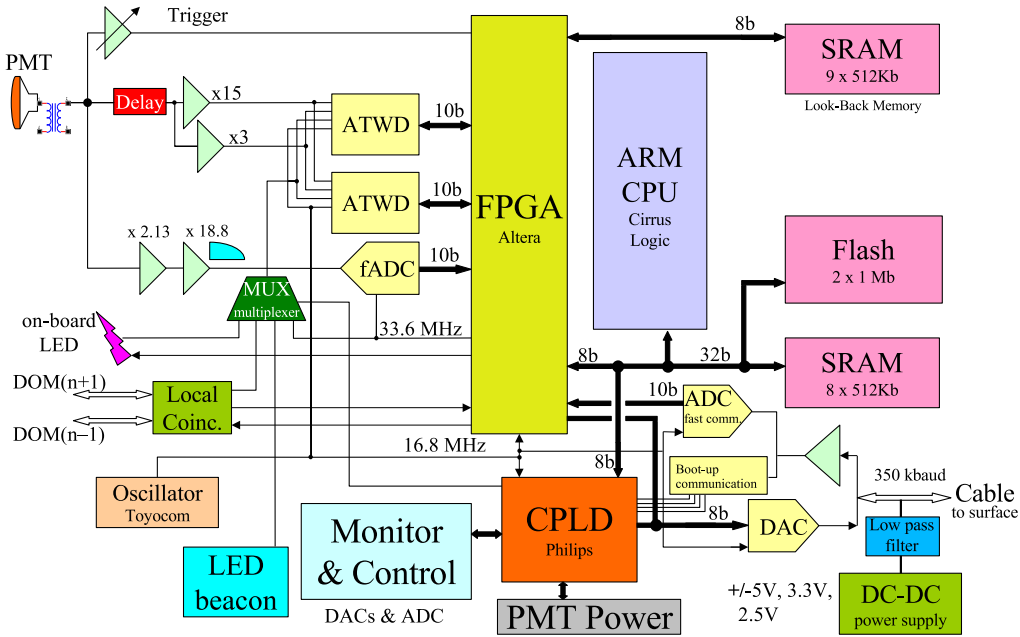}
\caption{Block diagram of the digital optical module signal processing
circuitry.} 
\label{fig:DOM_Circuitry}
\end{figure*}
Fig.~\ref{fig:DOM_Circuitry} shows the principal components of the
Digital Optical Module signal processing circuitry:  the
analog transient waveform digitizer (ATWD), a low-power custom
integrated circuit that captures the waveform in 128 samples at a rate\footnotemark
of $\sim$~600~MHz; a fast ADC operating
at\addtocounter{footnote}{-1}\footnote{The sampling speeds
in \ice are different from those chosen for \st.} $\sim$~30 MHz covering
several microseconds; a FPGA that provides state control, time stamps
events, handles communications, etc.;  a low-power 32-bit ARM CPU with
a real-time operating system. A very stable ($\delta f/f \sim 5 \cdot
10^{-11}$ over $\sim 5$~s) 
Toyocom 16.8 MHz oscillator is used to 
provide clock signals to several components.
Short cables connecting
adjacent DOM modules enable a local time coincidence (see Sec.~\ref{sec:LocalCoin}), which eliminates
most of the $\sim$1~kHz of noise pulses when enabled.

The DOM is operated as a slave to the surface DAQ. Upon power-up, it
enters a wait state for an interval of a few seconds, to permit
downloading of new firmware or software. If no messages are received
before time-out, the DOM will boot from Flash memory and await
commands. 

Normal operation of the DOM is dominated by a data-taking
state (see Sec.~\ref{sec:info}). In addition, a local-time calibration
process is periodically invoked, so that the local DOM time may be
ultimately transformed to master clock time with nanosecond accuracy
(see Sec.~\ref{sec:RAPCal}). The time calibration process appears as a
separately scheduled thread, with no deleterious impact upon the
data-taking state. Other active threads are message receive/send and
data transmission, also invisible to the data-taking process. There
are several calibration modes, such as running optical beacons, during
which normal data-taking of the respective DOM is inactive.  

Low power consumption was a high priority for the development of the
DOM circuitry because electrical power is limited at the South Pole.
Each \st prototype DOM uses only 3.5~W when booted
up and transmitting data.  An important feature in this respect is the
power consumption specific to the PMT pulse capture process since
electronic devices operated at high frequencies usually impose
significant demands.

\paragraph{PMT pulse capture}
\label{sec:info}
The DOM is equipped with innovative circuitry that is well-matched to
the PMT pulse characteristics and dynamic range. 
The DOM concept takes advantage of the fact that, most of the time,
no PMT pulses are present: The average time between pulses is about
$2$~ms. Circuit activity need occur only when pulses appear. The
waveform (pulse shape) capture capability is realized through a custom
switched capacitor array based 
Application Specific Integrated Circuit (ASIC) 
designed at Lawrence Berkeley National Laboratory (LBNL), the Analog
Transient Waveform Digitizer (ATWD)~\cite{Kleinfelder:2003}. The DOM
has two such ATWDs with identical tasks for fast switching from
one digitizing ATWD to the other ATWD.
This minimizes the dead time due to the digitization process.

The ATWD
has four channels, each with 128 samples, 
that synchronously record different input waveforms. Its employment in
the DOM is particularly advantageous because its power consumption is
only 200~mW. The ATWD performance combination of up to a GHz
sampling speed with very low power dissipation is unmatched
by any single commercial device for waveform capture. 
The DOM prototypes in \st are equipped with a $15\times$ and a
$3\times$ gain channel (see Fig.~\ref{fig:DOM_Circuitry}). The third channel is routed to an analog
multiplexer (MUX) that can connect to various internal signals for
diagnostic purposes. The 
fourth ATWD channel in \st DOMs is permanently connected to the
DOM internal clock, for calibration and monitoring ofthe ATWD sampling speed. 

The upper limit of linearity in pulse height for the lower gain
channel of the ATWD was found to be about
150~photoelectrons for a single pulse. This was achieved without
extensive optimization of PMT voltages and threshold.

\begin{figure}
\includegraphics[width=\columnwidth]{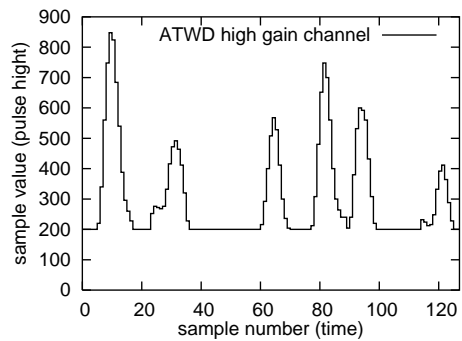}
\caption{Waveform likely caused by a down-going muon recorded in the DOM
closest to a reconstructed track.}
\label{fig:waveform}
\end{figure}
The waveform capture process is initiated by a
``launch'' signal derived from a discriminator connected to the
signal path. The capture process stores waveform samples as analog
voltages on four internal linear capacitor arrays. A single DC current
applied to the ATWD
controls the sampling speed. For \st we have chosen to capture the
pulse shape at 600~MHz (1.7~ns/\mbox{sample}). This high speed capture
is useful for characterizing PMT signal shape with good resolution and
for other diagnostic purposes. 
All four channels within an ATWD are sampled syncronously, with
aperture variations between channels measured to 
be less than 10~ps. Fig.~\ref{fig:waveform} shows a particularly
feature-rich example of a waveform as recorded by an ATWD with its
high-gain channel.
The data were obtained with zero suppression 
and data compression (see Sec.~\ref{sec:network}) in place. Due to the zero
suppression all values below the threshold of 200\footnote{corresponding to a
quarter of the average single photoelectron amplitude with respect to
a baseline level of 100.} are represented as
the threshold value.
The muon track reconstruction discussed later in
Sec.~\ref{sec:reco} indicates that the PMT signal was caused by a down
going muon with this DOM being the closest to the muon track. 
This plot demonstrates the capability of separating individual
pulses and other details within the ATWD waveform.
Recording the full structure of waveforms will aid in 
reconstructing complex events.

For waveforms that exceed the ATWD capture interval, sampling at a lower
frequency is sufficient to capture relevant information at later
times. For this purpose, the DOM is equipped with a 10-bit low-power
fADC\footnote{This fADC is a fast, 10-bit, parallel-output, pipeline
ADC}, operating at the DOM local clock frequency of 33.6~MHz. The PMT
signal in this path is stretched to match the lower
sampling rate.  

\paragraph{Local time}
Each PMT hit must be time-stamped such that correlated hits 
throughout the detector 
meet the relative timing accuracy requirement of 5~ns
rms. This is accomplished within the DOM using a two-stage method
that produces a coarse time-stamp and a fine time-stamp.
The \st DOM FPGA maintains a 54-bit local clock running at the DOM
33.6~MHz clock frequency.  
Once data reach the surface, the local time units as defined by the
local clock reference are transformed to master clock units, and are
ultimately linked to GPS (see Sec.~\ref{sec:RAPCal}). 
The 33.6~MHz is obtained by frequency-doubling the free-running Toyocom
16.8~MHz quartz oscillator. 

The PMT signal triggers a discriminator. The discriminator pulse is then
resynchronized to the next DOM clock transition edge providing the
coarse time-stamp. At 33.6 MHz, this coarse time-stamp has a time
quantization of 29.8~ns.
The resynchronized pulse of the discriminator provides the ATWD
``launch" signal.  The synchronous launching causes the PMT signal to
arrive within a one-clock-cycle wide time window at the ATWD. 
The leading edge\footnote{More sophisticated methods than the leading edge to
determine the fine time-stamp can be easily envisaged.} of the PMT
signal waveform within the ATWD record will occur somewhere within
this particular 29.8~ns interval. This is the ``fine time-stamp''
which provides the desired time resolution of less than 5~ns.  

For this to work, the PMT signal must be delayed sufficiently such that
the leading edge of the signal always appears well after the ATWD
sampling action has begun. An overall delay 
of about 75~ns is needed to accommodate both the coarse time-stamp
random delay and the propagation delay of the trigger processing path
the circuitry. In the \st DOM prototypes, a coil of coaxial cable of
15~m length was employed for this purpose.

\subsection{Network: Data link to the surface}
\label{sec:network}
All timing calibration, control, and communications signals between a
given DOM and the surface DAQ are provided by one conventional
twisted pair of 0.9~mm diameter copper conductors. This pair also
supplies power to the DOM. The resulting cable 
link is both cost-effective and robust, since the number of connections 
needed for all functionality is minimal. The twisted pairs are
assembled first in double pairs, as ``twisted quads'', and enclosed in a
common sheath. The required number of twisted quads per string are
then wrapped around a Kevlar strength member and covered with a protective
sleeve. Breakouts of individual quads to facilitate connections to the
DOMs are introduced at the appropriate positions.

The gross bandwidth across this 2.5~km long copper cable reached in
\st was about 35~kilobytes/s.
With two ATWD waveform channels of 128 10-bit sample values each and
an fADC of 256~10-bit samples the data size of a hit is 640 bytes not
accounting for overhead. The hit rate in \st is about 500~Hz
while the single pulse rate is about 1~kHz.\footnote{The 
discrepancy of the pulse and hit noise rates is caused by the significant
correlation of the noise - i.e. a single hit on average contains two pulses.
The noise is mainly caused by radioactive decays in the glass of the pressure
spheres and subsequent luminescence. The slow decay of the
luminescence gives rise to the intrinsic correlation of the noise.}
The resulting data rate is in excess of 300~kB/s which clearly exceeds the
available bandwidth.

We have developed two methods to reduce the amount of data to be
transfered from the DOM to the surface: local coincidence and
data compression.

\subsubsection{Local coincidence}
\label{sec:LocalCoin}
The DOM noise rate is approximately 100~times
greater than the rate induced by high-energy muons. Unlike hits
due to high energy particles, noise hits in one DOM occur
uncorrelated with hits of neighboring DOMs, i.e. primarily as
geometrically isolated hits. A local coincidence requirement,
imposed in the ice, selects predominantly only interesting hits for
transmission to the surface. Each DOM communicates with its nearest
neighbors by means of a dedicated short copper wire pair. 
The DOM is capable of sending and 
receiving these short signals in a full duplex mode across these
wires. Once a DOM has triggered an ATWD, it sends signals to each of its
neighbors. Meanwhile, a DOM is receptive to pulses from either or both
of its neighbors.  In local coincidence mode, only when this pulse
arrives, will the DOM digitize, store, and subsequently transmit the
time-stamped data to the surface.  

\subsubsection{Data compression}
Since the local coincidence requirement eliminates a small fraction of
real events~\cite{LocalCoin:1998}, an alternative compression of all
data including 
the noise hits has been developed.
The waveforms captured by a single ATWD channel or the fADC are
similar to a facsimile scan line. 
The original facsimile encoding standard~\cite{faxen}  is based on ``run-length''
encoding followed by Huffman encoding~\cite{huffman} with a fixed,
immutable, Huffman code. 
The adaptation of this idea to the digital
waveform data has led us to a 4~step algorithm:
\begin{description}
\item[I: Gain channel selection:]
In most cases a waveform does not saturate or exceed the
dynamic range of the high-gain channel 
of the ATWD. In these cases the low-gain channel information
is redundant and discarded. However, should the high-gain channel
saturate 
this channel is discarded and
instead the low gain channel data is preserved.
\item[II: Zero suppression:]
The main purpose of this step is to reduce the entropy of the data set
which makes the subsequently applied compression algorithms work more
efficient. A threshold of a quarter  of the average single
photoelectron pulse height is applied to the data of the highest
gain channel. This equivalent of a discriminator
threshold is then subtracted from all sample values. Resulting
negative values are set to zero.   
\item[III: Run-length encoding:]
This compression step replaces sequences (``runs'') of
consecutively repeated values with a single number prefixed by the
length of the run.  
In the standard definition of run-length encoding the prefix count
denotes occurrences and hence is always non-zero. In order to further
gain efficiency for the compression in the subsequent {\em
Huffman-lite} algorithm we choose to take the number of consecutive
repetitions instead i.e. the prefix count is one less than with the
standard definition and sample values occurring only once are prefixed
with a zero. As a consequence about 50\% of the resulting code are zeros.  
\item[IV: Huffman-lite encoding:]
Huffman coding simply uses shorter bit patterns as a replacement for
more frequent values in the input stream than for rarely occurring
values. We have reduced this concept to {\em Huffman-lite} in order to
save computing resources. The bits used for zeros are
minimized while we make no attempt to compress finite values.
We convert a zero value ``0'' to a single bit '0'. In order to
identify finite values these are preceded by a single high bit '1'. 
\end{description}
For the \st configuration this compression reduces the data portion to
15 bytes per hit on average which is a factor of 40 down from the
uncompressed data volume.
In \st this compression algorithm is executed by the CPU in the DOM.

The \st data presented in Secs.~\ref{sec:IceStud} and~\ref{sec:muons} have
been obtained using the compression to save bandwidth but 
without
the local coincidence criterion to maximize the number of contributing
DOMs and for straightforward 
comparison with the simulation based on the \aman detector and its
trigger logic (see Sec.~\ref{sec:CompSim}).

\subsection{Surface DAQ}
At the surface, the copper cables leading up from the DOMs are
connected to custom communications cards (\dc cards) in five
industrial PCs with a Linux operating system.  
Aside from the \dc cards all components of the surface DAQ are
commercial hardware.
The \dc cards can be
thought of as specialized serial ports for communicating with the DOMs
with additional functions for powering on and off each DOM, and
sending and receiving time calibration pulses.  
The five DOMCom PCs and a sixth, master control PC are connected by an
Ethernet network.
This network is then accessible via the South Pole
station LAN and, at certain times during the day, to the outside world
via satellite connection. 
The principal task of the surface DAQ is to allow one to take control
of and communicate with the DOMs in the ice for calibration and monitoring
purposes and to collect data from the optical sensors.   

FPGA logic in the \dc boards provides input and output FIFOs connected
to UARTs\footnote{A UART or universal asynchronous
receiver-transmitter translates between parallel bits of data and
serial bits.}  in the \dc boards.  The UARTs
in turn drive the 
communications circuits. 
Byte-wise communications of the \dc interface to the DOMs is provided
by a Linux kernel module device driver. By writing to and reading from
FPGA registers mapped to port addresses the kernel driver allows Linux
programs to communicate with the DOMs.

A suite of DAQ programs has been written to provide user interfaces
for monitoring, control (HV settings, etc.) and data-taking
purposes. Also, tools are available to reconfigure \st in many
respects.
Almost all of the components in the DOM can be reprogrammed after it
is deployed
in the ice.  These include the DOM boot PROM, the code executed by its
CPUs, and the FPGAs of both the DOM and the DOMCom card. 
For example, high speed communication and
data compression in the DOM were implemented after the deployment and
substantially improved the performance for physics data-taking.

\section{Local to global time transformation}
\label{sec:RAPCal}
\begin{figure}
\includegraphics*[bb=15 18 500 518,width=\columnwidth]{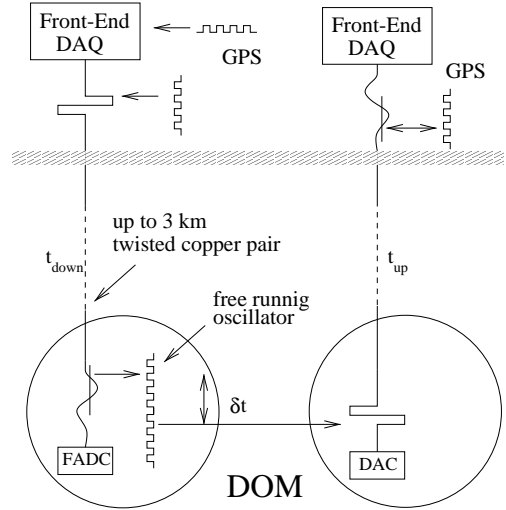}
\caption{General concept of the Reciprocal Active Pulsing (RAP) method
for time calibration.} 
\label{fig:rap}
\end{figure}

The (local) time stamp for a
photon in a DOM must be converted to a global, i.e.,  array-wide time
base, in order to be compared with hits in other DOMs.  Correlation
with events in nearby detectors, e.g. \aman and \textsc{IceTop}, or
much more distant detectors (satellites, other telescopes)
requires converting the global time to universal time.
Therefore, a critical requirement is the ability to calibrate the DOM
local oscillator against a master clock at the surface.

\subsection{Global time distribution}
A GPS clock at the South Pole represents this master-clock for \st. Its
10 MHz output is coupled to a special phase-locked loop on a clock
distribution circuit.  This PLL was set to give a frequency of
16.8~MHz in order to match the frequency of the local oscillators in
the DOMs. The 16.8 MHz signal is fed to each \dc board in the five \dc 
PCs where an additional PLL produces both 16.8 and 33.6 MHz phase-locked
signals. The time jitter introduced by the PLL distribution system
was measured to be less than 1~ns. 

\subsection{Reciprocal Active Pulsing}
Timing pulses sent 
down the string to a DOM at known time intervals are used to determine
relative frequency.
In response to the calibration pulse received at the DOM, 
after a well-defined delay $\delta t$,  a bipolar
signal identical to the one sent down the string is sent back to the
surface. With the pulses sent in both directions the offset of the two
clocks at the surface and in the DOM are determined.  
The left part of 
Fig.~\ref{fig:rap} 
depicts the transmission of the timing pulse from the surface down to
the DOM 
while the right hand side shows the transmission in the reverse
direction with the same DOM and \dc combination. It is important
that the circuitry for generation and recording of the timing pulse be
identical at the DOM and at the \dc card and that the treatment is the
same in both places. This way, the shapes of
the pulses sent down and up are identical and can be analyzed in the
same way to determine the time mark.  
This procedure yields an overall delay $t_\text{rt}$ for the full
round-trip. With the same signal treatment at the surface and in the DOM,
one gets 
\be
t_\text{up} = t_\text{down} = \frac{1}{2} \left( t_\text{rt} - \delta
t \right) \  .
\label{eqn:trt}
\ee
This calibration method is called Reciprocal
Active Pulsing~(RAP)~\cite{RAP:1998}. It maps the
local clock counter values in the DOM and the hit time-stamps to the
global time.

\begin{figure}
\includegraphics[width=\columnwidth]{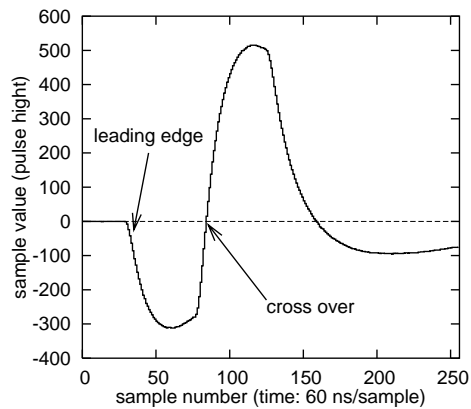}
\caption{The timing pulse emitted by the \dc card recorded over a
15~$\upmu$s period as received by the DOM.}
\label{fig:RapDOM}
\end{figure}
Fig.~\ref{fig:RapDOM} shows the shape of the timing pulse observed
by the DOM at the end of a $\sim$2~km cable for a bipolar square-wave
pulse injected 
by the \dc card.  The pulse displays the 
effect of cable attenuation and dispersion.
The pulses
are recorded at surface and at depth by fADCs whose sampling rates
(16.8 MHz) are fixed by either the local oscillator in the DOM or the
master clock on surface. The shapes of the pulses received by the DOMs
and the \dc cards are similar and indeed allow for the identical
treatment.

Two methods were used to determine a reference time mark on the pulse:  
\begin{itemize}
\item The intersection of the baseline voltage (linear fit)
with the negative-going slope of the leading edge\footnote{Tangent to the
inflection point of a cubic function fit to the leading edge}.
\item  The intersection of the crossover point (linear fit) with the
extrapolated baseline.    
\end{itemize}

We found that leading edge timing is least sensitive to variations in
component values such as temperature variations after powering up. 
Crossover timing, however, gives better time resolution and has less
jitter from one time calibration cycle to another. 
It was found that temperature drifts that show up in the
cross over timing change the times $t_\text{up}$ and $t_\text{down}$
in identical amounts because the drifting electronic components affect both
the up-going and the down-going pulses in the same way. This implies
that the effect cancels for the 
calibration of the local time. Moreover, the observed drifts are very
slow, less than 15~ns in 1~hour. Therefore, we have used the crossover
timing to obtain the local to global time calibration.
\begin{figure}
\includegraphics[width=\columnwidth]{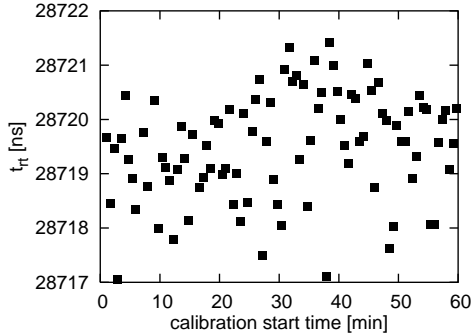}
\caption{Measured round-trip time $t_\text{rt}$ from successive RAP calibrations for
a DOM at 2~km depth.}
\label{fig:trt}
\end{figure}
Fig.~\ref{fig:trt} shows the measured round-trip time $t_\text{rt}$
(see Eq.~(\ref{eqn:trt})) over a period of one hour for a DOM at 2~km depth employing the cross
over timing.
The jitter is 1.1~ns rms, which is a relative error of only
$4\cdot 10^{-5}$. 

Performing these clock calibrations automatically every
10 seconds -- as was done during data-taking with \st~-- consumes
negligible bandwidth. We will demonstrate in Sec.~\ref{sec:timing}
that this RAP method at this repetition rate provides a timing
accuracy of better than 5~ns for the pulse times of a DOM.

\section{Studies with optical beacons}
\label{sec:IceStud}
To study the performance and uniformity of \st we have performed
various dedicated data-taking runs operating the optical beacons of
all available DOMs. 

\subsection{Timing accuracy}
\label{sec:timing}
In order to determine the timing accuracy of the RAP calibration
procedure described in Sec.~\ref{sec:RAPCal} we have observed the
light emitted by the LED beacon of one DOM by two neighboring DOMs.
The beacon is set to a very bright output operation mode. For short
distances the light front created at the beginning of a flash is to a
good approximation unaffected by scattering in the ice as photons
undergoing scattering fall behind the front. As long as there are
enough photons as part of the light front, this front moves with the
speed of light in the ice, $c/n$, where $n$ is the refractive index.
Hence, this light flash provides a reference time signal.

\begin{figure}[h]
\includegraphics[width=\columnwidth]{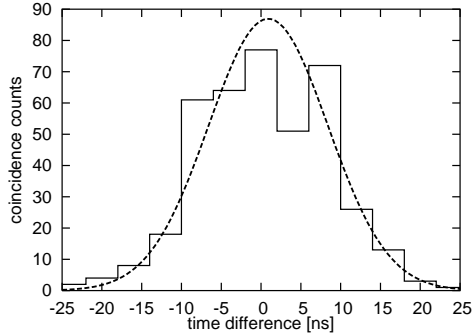}
\caption{Relative timing accuracy of a pair of DOMs measured with a
neighboring flashing beacon.}
\label{fig:timing}
\end{figure}

Fig.~\ref{fig:timing} shows the measured time difference of the arrival times
of this light front at both DOMs corrected for the propagation time of
the light from one DOM to the other. One observes a distribution with
a square root variance of 7.5~ns. The jitter includes effects of
the PMT timing and the RAP procedure. Accordingly, this provides an
upper limit for the timing accuracy of the RAP calibration
process. Since both DOMs contribute to the variance of the observed
time difference, the accuracy for an individual DOM with respect to the
global surface time is about 5~ns.

\subsection{Light propagation in the ice}
\label{sec:LightProp}
We have studied the propagation of the LED light from the optical
beacon (see Sec.~\ref{sec:beacon}) through the ice.
From the beginning of \aman operations, it has been crucial to
establish an understanding of the properties of the glacial ice at the
South Pole~\cite{Ice1000m}. 
Today the optical properties of the ice are
mapped over the full relevant wavelength and depth
range~\cite{Woschnagg:2004qq} using the analog optical modules. 

\begin{figure}
\includegraphics[width=\columnwidth]{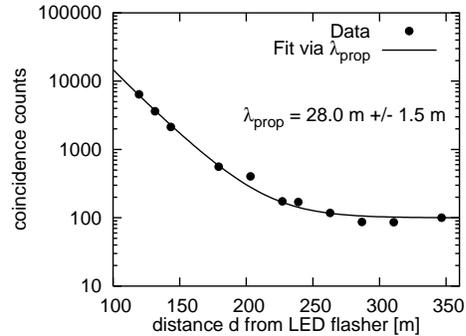}
\caption{Attenuation of light from the optical beacon as a function of
the distance from the source.}
\label{fig:absorption}
\end{figure} 
Using a configuration similar to the one discussed in
Sec.~\ref{sec:timing} we have used the flashing beacon in a DOM and
observed the light in two other DOMs, one being the next neighbor to
it. Here, the other 
observing DOM can be any other one along the
string. Fig.~\ref{fig:absorption} shows the coincidence counts of the
two observing DOMs as a function of the distance $d$ of the second
observer DOM from a flasher at a
depth of 1993~m below the ice surface. 
The first observing DOM as a neighbor
to the beacon is close enough to observe every emitted flash. For
DOMs further away photon counting in coincidence with the neighbor DOM
is possible. 
The combined effect of scattering
and absorption in the ice leads to a fall off of
the light intensity $I$ with an effective propagation length parameter
$\lambda_\text{prop}$: $I(d) \propto
\exp(-d/\lambda_\text{prop})/d$~\cite{Askebjer:1997}. Random coincidences of
the two observing DOMs due to the dark noise of the distant
DOM\footnote{The dark noise of the neighbor DOM can be neglected compared
to the high flash rate.} yields an additional offset $c_2$ independent of the
distance. From fitting $c_1 \exp(-d/\lambda_\text{prop})/d + c_2$ via
$c_1,\ c_2$ and $\lambda_\text{prop}$ we obtain
$\lambda_\text{prop} = 28.0~\pm~1.5\,\text{m}$. With the described
setup $\lambda_\text{prop}$ represents a measurement of the average
propagation length at depths between 1750~m and 1880~m. This is in
excellent agreement with previous findings with the analog
technology~\cite{Woschnagg:2004qq}.   

\begin{figure}
\includegraphics[width=\columnwidth]{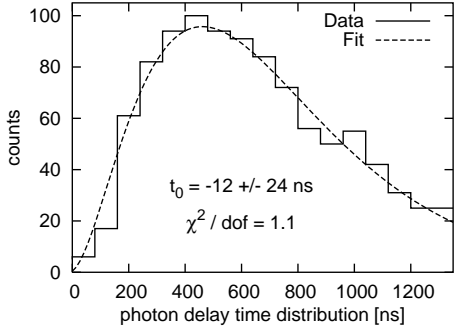}
\caption{LED photon arrival time distribution due to scattering
through 143~m of ice}
\label{fig:scattering}
\end{figure}
With the same setup we have also observed the arrival time distribution of
the photons at the distant DOM (see Fig.~\ref{fig:scattering}). Here, the
DOM next to the beacon 
provides the reference for the time of the
flash. The time distribution is shifted to
account for the time it takes light to traverse the ice without
scattering from the beacon to the distant DOM (delay time
distribution). The arrival time distribution 
for the DOM next to the beacon can be neglected as discussed in
Sec.~\ref{sec:timing}. Fitted to the data is a gamma distribution 
\par\nobreak\noindent\begin{equation}
f(t|t_0,\alpha,\beta) = (t-t_0)^{\alpha-1} \frac{\beta^\alpha
e^{-\beta (t-t_0)}}{\Gamma(\alpha)} 
\label{eqn:gamma}
\end{equation}\par\nobreak\noindent
for $t-t_0 > 0 $ with a shape parameter $\alpha>0$ and a scale parameter
$\beta>0$.  
Such a gamma function was found to describe the arrival time
distributions reasonably well with scattering and
absorption in ice or water for the case of an isotropic,
monochromatic and point-like light source
~\cite{Pandel:1996,Ahrens:2003fg}.  
The observed time distribution with the DOM follows this known
behavior. As an additional parameter to the usual definition of a gamma
distribution in Eq.~(\ref{eqn:gamma}) we have introduced $t_0$ to
account for possible timing errors of the DOM. It turns out that $t_0$
is compatible with zero which is consistent with the results presented
in Sec.~\ref{sec:timing}.

\section{Performance and analysis with atmospheric muons}
\label{sec:muons}
Atmospheric muons provide a test beam for neutrino
telescopes, as was demonstrated by \aman~\cite{Andres:2000}. We will
use it here to demonstrate the performance of \st.

\begin{figure*}
\includegraphics[width=0.5\textwidth]{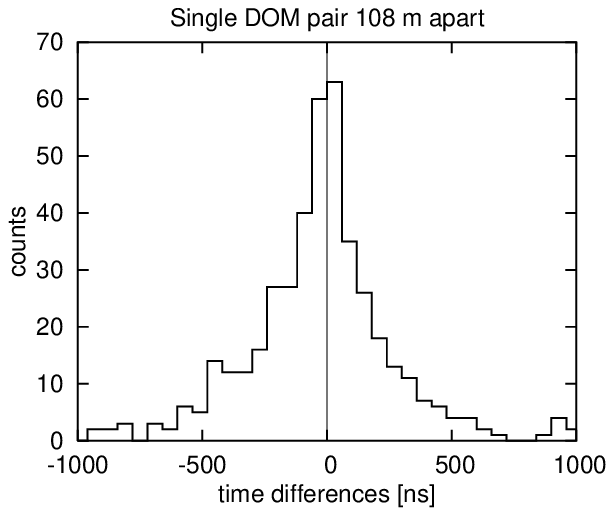}
\includegraphics[width=0.5\textwidth]{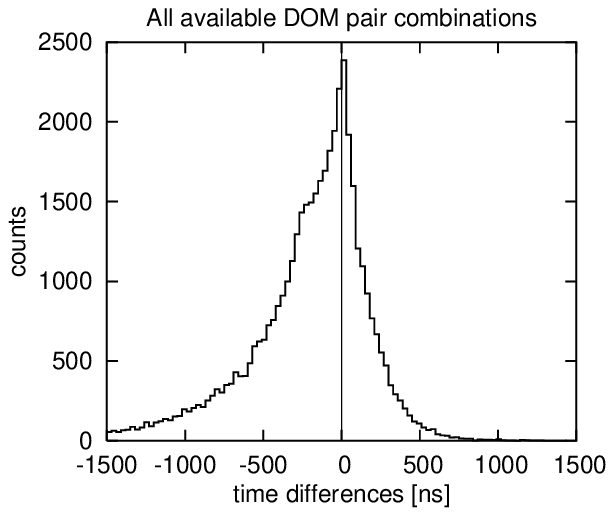}
\caption{Time differences between pairs of DOMs corrected for the
down-going muon hypothesis. Left: single DOM pair 108~m apart, right:
all available DOM pair combinations accumulated.}
\label{fig:DOMpairs}
\end{figure*}

\subsection{Down-going muons}
The vast majority of the atmospheric muons that reach the detector
travel at near the vacuum speed of light. The zenith angle
distribution is peaked in the downward direction. We have compared the
coincidence hit times of the DOMs with the hypothesis of a straight
down going muon, i.e. we have subtracted the time it takes the muon
from the upper DOM to the lower one from the hit time
differences. Fig.~\ref{fig:DOMpairs} shows the distribution of these
time differences under the condition that at least 4~DOMs
of the string registered the event. On the
left of Fig.~\ref{fig:DOMpairs} we show the distribution of
time differences 
of a particular pair of DOMs. The two DOMs are 108~m apart. The right
hand side of Fig.~\ref{fig:DOMpairs} shows an accumulation of these 
diagrams for all available pairs of DOMs. 
With the hypothesis of a straight down-going muon, the peak count
should be at zero. 
The exact location of the peak at zero to within a
nanosecond, which can be seen by examining the peak region in more
detail than shown, is another evidence for the accurate timing and the
proper functioning of the RAP calibration method.  

The asymmetry of the tails of the
distribution towards negative time differences is a consequence of not
quite straight down-going muons. For small zenith angles of the muon
track the \cer 
light that reaches the DOMs does indeed cause hit times that proceed 
from top to the bottom of the string that appear faster than the
vacuum speed of 
light. 
A muon track at zenith angle $90^\circ - \theta_\text{\v{C}}$
($\theta_\text{\v{C}}$ is the Cherenkov light emission angle) that
intersects the string will cause all DOMs located above the
intersection point to light up at once (neglecting scattering in the ice). 
By shifting the time differences with the
hypothesis of a straight down going muon one obtains these negative
time differences.  We will see in Sec.~\ref{sec:reco} that these
deviations can indeed be used to reconstruct the zenith angle of the
muon track. 

\begin{figure}
\includegraphics[width=\columnwidth]{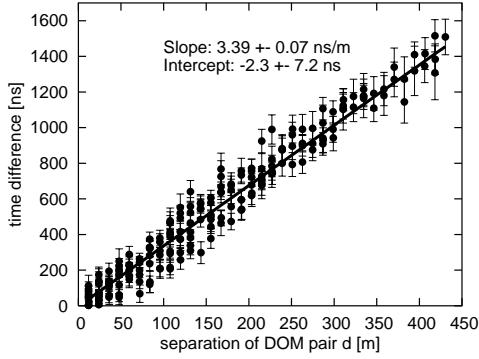}
\caption{Time differences between all available pair combinations of
DOMs as a function of the separation.}
\label{fig:MuonGeometry}
\end{figure}

Fig.~\ref{fig:MuonGeometry} shows the mean time differences of all
available pairs of DOMs as a function of the separation of the
DOMs. For this plot the time differences were not 
shifted for the down-going muon hypothesis and we require that
no fewer than 6~DOMs have ``seen'' the muon. A linear fit to the data shows
a slope compatible with straight down-going muons traveling at the
speed of light. The intercept of the fit at
$-2.3\,\text{ns}\pm7.2\,\text{ns}$ is another indication that the
local to global time transformation works.

\subsection{Zenith angle reconstruction}
\label{sec:reco}
In the absence of scattering and absorption in the \cer medium
the arrival time of the \cer light from the muon at the string
takes the simple analytic form of a hyperbola as a function of the depth
$z$ along the string~\cite{Carmona:2002xw}\footnote{We have to 
caution that the approximation presented as Eq.~(5) in
Ref.~\cite{Carmona:2002xw} is wrong for all $z-z_0<0$. We have not
used this approximation but the exact description of the hyperbola
with our Eq.~(\ref{eqn:hyp}).}: 
\par\nobreak\noindent\begin{equation}\begin{split}
t = & \ t_0 + \frac{1}{c} \left\{(z-z_0) \cos \theta + 
\phantom{\sqrt{\sin^2 \theta}}\right.\\
 & \left.\sqrt{(n^2-1) [d^2 + (z-z_0)^2 \sin^2 \theta] } \right\}
\end{split} \label{eqn:hyp}
\end{equation}\par\nobreak\noindent
Here, $d$ is the minimum distance between the string and the muon
track; $\theta$ is the zenith angle of the muon;
$t_0$ and $z_0$ are the time and the depth at which this closest
approach of the muon occurs. 
Since $t_0$ is not directly related to
the observable hit times of the DOMs we replace it by $t_\text{first}
= t_0 + d\sqrt{n^2-1}/c$ which is an estimate for the time when the
\cer light from the point $(t_0,z_0)$ reaches the
string\footnote{This of course disregards the initial orientation of
the \cer light. Scattering in the ice however largely voids this
feature.}. 

The 4 parameters $t_\text{first}, z_0, d$ and $\theta$ are to be
fitted to the data. We have used the {\em GNU Scientific
Library}~\cite{gsl} to perform this nonlinear multiparameter fit. 
Computing the Jacobian matrix from explicit
derivatives greatly improves the numerical stability and provides fit
results practically independent of the initial guess of the parameters
as long as they are chosen within a physically reasonable region --
e.g. $z_0$ in between the actually hit DOMs. To calculate the
derivatives we have used the computer algebra system {\em Maxima}~\cite{Maxima}.
For the initial guess we have used $d=0$ and $\theta = 90^\circ$. The
guess for $z_0$ was computed as the average $z$ of all hit DOMs
weighted with the pulse charges. The initial value of $t_0$
was set to the hit time of the DOM next to $z_0$.
Eq.~(\ref{eqn:hyp}) holds for the idealized situation without
scattering and absorption in the medium. The ice at the
South Pole however shows scattering and absorption.
By integrating the gamma function mentioned in
Sec.~\ref{sec:LightProp} with the empirical parameters successfully
used in \aman~\cite{Ahrens:2003fg} one obtains that the mean speed of
propagation of the light is about 11~ns$/$m in contrast to the
expectation of about 5~ns$/$m based on the refractive index of ice
of $n = 1.32$. 
To account for that we have replaced $n$ in
Eq.~(\ref{eqn:hyp}) by $n_\text{eff} \simeq 3.16$. Note that
the effective opening angle of the light cone\footnote{The
initial emission angle of the \cer light is unaffected, of course.} of
the produced \cer light is reduced as well. This is why a simple
replacement of $n$ by $n_\text{eff}$ in Eq.~(\ref{eqn:hyp}) is
consistent. 

\begin{figure}
\includegraphics[width=\columnwidth]{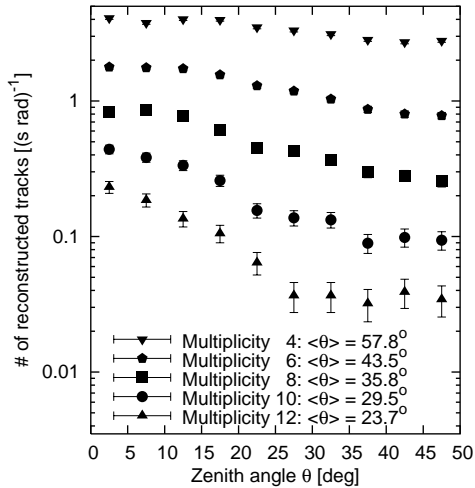}
\caption{Reconstructed zenith angle distribution with \st for several multiplicity
conditions}
\label{fig:angles}
\end{figure}

Fig.~\ref{fig:angles} shows the zenith angle distribution as
reconstructed with this method applied to \st. Here, a zenith angle of zero
denotes a straight down-going muon.  
One
observes that the angular distribution falls off faster with higher
required multiplicities of the number of hit DOMs. This is an effect
related to the acceptance changing with increasing multiplicity. Due to
the absorption of the \cer light in the ice, only close by
vertical tracks have a chance to be seen by all DOMs. However close a
track might be, even when intersecting with the string, a horizontal
track has a very low chance to light up more than only a handful of
DOMs.

\subsection{Comparison with simulation}
\label{sec:CompSim}
We have compared the observed rates of muon tracks with a detailed
simulation in order to verify that \st indeed performs with a homogeneous
efficiency. The acceptance for muon tracks agrees with the
expectations based on the parameters of the whole light collection
process i.e. the optical coupling of the ice to the pressure sphere, the
glass and gel transmission and the quantum efficiency of the PMT.
The simulation is based on the air shower generator
\textsc{Corsika}~\cite{Knapp:1998ra}, a program package for muon
propagation through matter MMC~\cite{DimaPhD,Chirkin:2004hz} and the
\aman detector simulation program
\textsc{Amasim}~\cite{Hundertmark:1998vw}. This simulation for the
full \aman detector was restricted to the \st geometry.

Besides generating hit times for the DOMs the simulation also provides
the zenith angle of the generated muon tracks. By applying the
reconstruction method of Sec.~\ref{sec:reco} to the simulated hits one
can study the performance of the reconstruction algorithm. It turns
out that the distribution of the deviation of the reconstructed zenith
angle from the generated muon track angle roughly follows an
exponential behavior $\propto \exp(-\Delta\theta/b)$ with $b \simeq
22^\circ$. This corresponds to a median error of $15^\circ$. 

This reconstruction technique is of course inferior to the full AMANDA
reconstruction~\cite{Ahrens:2003fg};
the main limitation here stems from the reduced geometry of
a single string versus an array of optical modules. Also the
chosen approach to reconstruction is much simpler. It was not the
primary aim of this 
analysis to provide an optimized reconstruction method. However, the
error of the reconstruction method is small enough to calculate
angular distributions that show the characteristic features.

\begin{figure}
\includegraphics[width=\columnwidth]{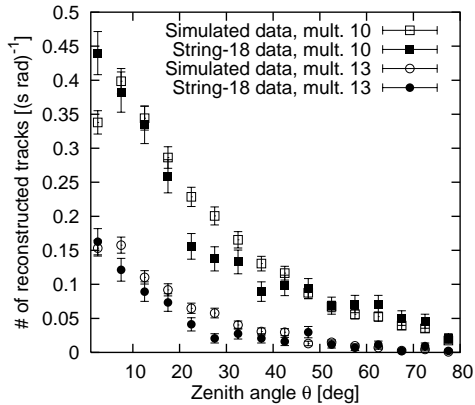}
\caption{Muon tracks reconstructed with \st data and with simulated
data for two multiplicity conditions.}
\label{fig:AngCompa}
\end{figure}

Fig.~\ref{fig:AngCompa} shows a comparison of muon tracks
reconstructed with \st data to tracks reconstructed with simulated
data. Note that the distributions have {\em not} been scaled with
respect to absolute numbers. The agreement found both in shape and in
absolute rates is excellent and
shows that efficiencies and timings are well under control in \st.

\section{Conclusion}
\st is established as proof of concept for the \ice array. The
all-digital technology with pulse digitization already at the PMT
inside the optical module, 
digital data transmission to the surface and automated time
calibration was shown to be fully functional including the benchmark
observation of atmospheric muons.

\begin{ack}
This research was supported by the following agencies: 
Deutsche Forschungsgemeinschaft (DFG);
German Ministry for Education and Research; 
Knut and Alice Wallenberg Foundation, Sweden; 
Swedish Research Council; 
Swedish Natural Science Research Council; 
Fund for Scientific Research (FNRS-FWO), Flanders
Institute to encourage scientific and technological research in
industry (IWT), and Belgian Federal Office for Scientific, Technical
and Cultural affairs (OSTC), Belgium.
UC-Irvine AENEAS Supercomputer Facility; 
University of Wisconsin Alumni Research Foundation; 
U.S. National Science Foundation, Office of Polar Programs; 
U.S. National Science Foundation, Physics Division; 
U.S. Department of Energy; 
D.F. Cowen acknowledges the support of the NSF CAREER program.
I. Taboada acknowledges the  support of FVPI.
M.~Ribordy acknowledges the support of the Swiss National Science 
Foundation.
K.~Helbing acknowledges the  support of the Alexander von Humboldt Foundation.
\end{ack}

\end{document}